\documentclass[10pt,twocolumn]{article}
\usepackage{graphicx}
\usepackage{amsmath}   
\usepackage{url}
\usepackage{multirow}
\usepackage{setspace}
\usepackage{ftnright}
\usepackage{wide}
\newcommand{\be}{\begin{equation}}
\newcommand{\ee}{\end{equation}}
\begin{document}
\title{Congestion Reduction Using Ad-Hoc Message Dissemination in Vehicular Networks} 
\author{Thomas Hewer$^{1,2}$ and Maziar Nekovee$^{1,3}$\\
\small$^1$ Department of Computer Science, University College London, Gower Street, London WC1E 6BT, UK \\
\small$^2$ Centre for Computational Science, University College London, 20 Gordon Street, London WC1H 0AJ, UK \\
\small$^3$ BT Research, Polaris 134, Adastral Park, Martlesham, Suffolk IP5 3RE, UK \\
t.hewer@cs.ucl.ac.uk, maziar.nekovee@bt.com}
\maketitle

\begin{abstract}
Vehicle-to-vehicle communications can be used effectively for intelligent transport systems (ITS) and location-aware services  \cite{Torrent-Moreno2007}.  The ability to disseminate information in an ad-hoc fashion allows pertinent information to propagate faster through the network.  In the realm of ITS, the ability to spread warning information faster and further is of great advantage to the receivers of this information.  In this paper we propose and present a message-dissemination procedure that uses vehicular wireless protocols for influencing traffic flow, reducing congestion in road networks.  The computational experiments presented in this paper show how an intelligent driver model (IDM) and car-following model can be adapted to `react' to the reception of information. This model also presents the advantages of coupling together traffic modelling tools and network simulation tools.
\end{abstract}

\addvspace{10pt}

\section{Introduction}

In the realm of vehicle to vehicle communications there are several methods for the dissemination of data that are being actively researched.  The use of satellite communication, such as those linked to global positioning services (GPS) offer global communication but require expensive equipment, large antennae and, due to the large distances the signal must travel, have a high latency.  Cellular telephone networks offer a lower latency but are still slow when communicating with other vehicles nearby, and require cellular contracts to use the network. The scenarios presented mainly in this paper require high-speed communication that disseminates from source, which is difficult to achieve using either cellular or satellite communications.

Ad-hoc networks offer a good method to spread information outwards from an origin quickly and efficiently.  It has been shown in \cite{1682943} that in ad-hoc networks worms spread in a epidemic pattern that can be modelled.  Using such modelling techniques we can develop algorithms that allow for a change to be made to the speed, position and route of a vehicle.  A further advantage of ad-hoc networking is the unlicensed use of the radio-spectrum and the recent reduction in cost for the equipment for communication. A separate leg of the research being undertaken on wireless fidelity (under IEEE standard 802.11) has been developed in the past few years specifically for vehicular ad-hoc networks (VANETs).  The 802.11p WAVE standard specifies network protocols which address the difficulties associated with vehicular networks.  These difficulties involve short link time, delay tolerance and the inefficiency of wired-network paradigms, that have been pulled into the wireless standards.

The following simulation experiments and algorithms were developed with particular scenarios in mind.  These scenarios operate in a dual-carriageway environment with no junctions and an obstacle or danger that is present at some point in the field.  The simulations use both traffic modelling and message propagation to advise the vehicles of the obstacle at a greater distance than line-of-sight provides.  The results show that by spreading information quickly and efficiently through the network we can develop algorithms that reduce congestion and other traffic flow effects. By coupling the telecommunications and driver model in one tool we can perform these simulations in the same runtime. This method of simulation can also be useful for intelligent transport systems, where simulations run in parallel can sweep the parameter-space for the best outcome.

\section{Simulation System}

The simulation tool we use is adapted from the dynamic traffic simulator by Treiber et al. \cite{treiber-2000-62}. This tool uses a simple model of a two-lane roadway, but contains an advanced driver model and lane changing algorithm, MOBIL \cite{Treiber1999}.  Through the addition of telecommunications and adaptions to both the intelligent driver model (IDM) and lane changing model, based on reactions to information received from an obstacle or danger, we can show that more efficient information dissemination through a network can increase system throughput and also reduce stop-and-go traffic formations.

\subsection{Vehicular Modelling}

To accurately model traffic behaviour, there are several key components: a driver model to develop how real people will drive under certain circumstances, a lane changing model to make realistic decisions on when would be advantageous to change lane and a roadway with rules (i.e. drive on the left in the UK).

A car following algorithm will contain at least a desired velocity, a safe time separation when following other vehicles, an acceleration and a braking criteria \cite{treiber-2000-62}. At each simulation time step the acceleration is calculated for each vehicle. The parameters of these models can be changed to emulate more aggressive and more considerate drivers.

When modelling vehicular networks over large areas (i.e. metropolitan areas) the flow of traffic  on a single road can be seen to operate as an incompressible fluid (as later results will show) according to $Q=\rho V$, where $\rho$ is the average density of traffic (cars/km) and $V$ is the average velocity on the road (km/h) \cite{4212940}.  At microscopic levels of simulation (across any field size) each object is treated as a discrete event operating independently of all others, which greatly increases the computational requirements of the system, but which provides a more realistic and component based approach to modelling.

The IDM in the simulator follows the MOBIL model \cite{Treiber1999} which was developed by M. Treiber.  MOBIL operates as a car-following model such that the acceleration and braking are defined by the distance from the car in front.  The function of such an acceleration $\frac{dv}{dt}$ is as follows:

\begin{equation}
\label{mobil1}
\frac{dv}{dt}= a \left[ 1 - \left( \frac{v}{v_0} \right)^\delta - \left( \frac{s^*}{s}\right) ^2 \right] 
\end{equation}
where
\begin{equation}
\label{mobil2}
s^* = s_0 + \left( vT + \frac{v\Delta v}{2 \sqrt{ab}} \right)
\end{equation}

for acceleration on an open road $a$, velocity $v$, desired velocity $v_0$, distance $s$ to front vehicle, desired dynamic distance to front vehicle $s^*$, velocity difference $\delta$, a safe time delay between vehicles $T$, a comfortable braking value $b$ and a minimum distance between vehicles $s_0$.

Lane changing algorithms add a necessary level of complexity to the IDM.  In order to decide whether to change lane or not, the current acceleration must be calculated for the current lane and the acceleration in the new lane (with regards to the car behind and in front in the new lane).  If the acceleration in the new lane is greater than that in the current lane, there is an advantage to be gained by changing lane.  Many models, including those in the original simulator, include a bias in the model for particular lanes, which simulates the real scenario of slow lanes and fast lanes.

\subsection{MAC Layer Protocol}

In simulating wireless fidelity networks, the majority of simulations use the IEEE 802.11 protocols  \cite{4248378} , as this offers the best simulation of the MAC layer functionality.  The IEEE 802.11 MAC layer uses a distributed co-ordination function (DCF), which has been simulated in \cite{weinmiller96analyzing}, to ensure efficient communication on the medium, and implements controls to reduce collisions.  More recently the IEEE 802.11p standard has been tested in  \cite{Yin2003} specifically for inter-vehicular communications. This allows the foundation of underlying strengths in the 802.11 suite to to be enhanced for vehicular networks.

\label{mac}
The MAC layer in the simulator operates using an adapted version of IEEE 802.11 which removes the inter-frame spacing (IFS) model, enabling equal priority to all network traffic. Due to implementation, and the need for simplicity in the model, our implementation of 802.11 does not suspend the back-off timer when the medium is busy during that frame, as 802.11 does.

The network back-off when the medium is busy $X$ operates as follows \cite{4212940}:
\begin{equation}
X \in 2^n \times [B_{min}, B_{max}] 
\end{equation}
where $n$ is the number of times it has previously had to back off in succession. $B_{min}$ and $B_{max}$ are the minimum and maximum possible back off time, respectively. $B_{min}$ is often set at 0. The medium is defined busy if any car within the transmitters interference range, $R_i$, is currently broadcasting.

Every car within the transmission range, represented by $R_c$, (which is usually twice as small as the interference range) will receive the message with probability $\lambda$.

\subsection{Radiowave Propagation}

In the modified simulator the reception of messages is performed by a basic algorithm controlled by the simulation engine.  In advanced network simulators the realistic reception of messages depends on the signal strength at the receiver.  The basic propagation model is the Friis free-space calculation, which extends the ideal free-space propagation formula ($Pr \propto1/d^2$)  \cite{Struzak2006} where Pr is received power and d is the distance from transmitter) to incorporate the antenna gain(both transmit and receive).  The Friis model, however, will only hold true with a clear line-of-sight (LOS) between transmitter and receiver, and assumes no level of scattering by atmospheric particles (which becomes very apparent in urban and highway scenarios \cite{Laasonen2003}).  Friis operates as follows:

\begin{equation}
\label{friis}
P_r (d)=\frac{P_tG_tG_r\lambda^2}{(4\pi)^2 d^2 L} 
\end{equation}

where: Pr is the received power at distance d with respect to the antenna gain and height and the system loss L.

One method of altering the propagation of messages through the simulated network, is to change the transmitted power and therefore the transmission range.  By transmitting information further the message has a greater probability of retransmission in sparse networks and also a faster dissemination through the system \cite{Torrent-Moreno2007a}.  This approach has both advantages and disadvantages, mentioned in detail later. The main disadvantage of having a large transmission range is that information propagates further, eventually reaching a point where the information usefulness is low, and so is taking up bandwidth and time for almost redundant data.  In a system with a limited time and bandwidth this can cause localised problems, where more pertinent (i.e. geographically closer) information is lost to redundant data.

\section{Algorithms}

This section examines the algorithms used in the simulation. These algorithms form the basis for the work we present  here and have been designed specifically for vehicle-to-vehicle and VANET scenarios. 

\subsection{Epidemic Message Passing Algorithm}

The propagation of messages through a system requires an efficient delivery algorithm.  In our simulation we use a probabilistic information dissemination protocol, which is fully defined in  \cite{4212940}. We assume that all vehicles know their location (via GPS technology) and that each message contains information about it's location and time of creation. To ensure propagation does not extend to redundant distances from source each message has a Time-To-Live (TTL) setting.

The algorithm allows for a reasonable amount of retransmission and dissemination through the network and balances the relevance of the information with the distance from the source.  To this end information can disseminate quickly and efficiently and also reduce information spreading to vehicles who do not require the information (as discussed previously, this can cause more pertinent information to be lost).

The probability for rebroadcasting, $P$, as described by  \cite{4212940} is obtained from:

\begin{equation}
\label{probrebroad}
P = \left\{ 
\begin{array}{ccc}
1 & $if$ N_f $or$ N_b = 0 \\ \\ 1- exp \left( - \alpha \frac{N_f - N_b|}{N_f + N_b} \right) & $otherwise.$ 
\end{array} \right.
\end{equation}

where $N_f$ and $N_b$ are the number of times the car has received that particular message from front and from back, respectively and $\alpha$ is a protocol parameter, which controls redundant transmissions. In the case of directional message propagation Eq.\ref{probrebroad} is modiÞed such that if a message is propagating in either direction it is only kept alive by nodes near the head/tail of the group. In this case the rebroadcasting probability, $P$, is computed from:

\begin{equation}
\label{rebroadprob}
P = \left\{ 
\begin{array}{ccc}
1 & $if$ N_k = 0 \\ \\ 1- exp \left( - \alpha \frac{N_k}{N_k + N_{\vec{k}}} \right) & $otherwise.$ 
\end{array} \right.
\end{equation}

where $N_k$ is the count of messages received from the direction of message propagation (e.g. if $k$ is forward, $N_k$ is the count of messages received from vehicles infront), and $N_{\vec{k}}$ is the count of messages received from the opposite direction.

\subsection{Variable Speed Limit}
\label{vsl}

The initial algorithm we introduce is to reduce the value of the desired velocity ($v_0$ in Eq. 5) by a fixed amount when the vehicle has received the warning message.  This achieves an overall slowdown in the network which can reduce the time delay between free-flow and gridlock (where $v = 0$) at the obstacle.  This particular algorithm change has a transient effect on the network, such that the system will still become congested over time. The idea for this came from the London Orbital (M25) variable speed limit which operates on parts of the motorway.

The value we reduce $v_0$ by is of great importance. Initial tests showed that by reducing the desired velocity too much (i.e. a reduction of over 10ms$^{-1}$) when infected, the effect on the network was to cause congestion further back in the system, such that gridlock (i.e all vehicles in the field of simulation are static) occurs much sooner.  By reducing the value of $v_0$ by 2.7ms$^{-1}$ the network slowed well and the time delay between free-flow and gridlock was increased without causing congestion further back in the network.

An important algorithmic change is the return to the normal value of $v_0$ once the obstacle has been passed geographically, otherwise the recovery from the obstacle will take a greater amount of time. We did test a proportional change in the desired velocity as the obstacle was approached, but this provided little observable effect at great distances and a highly-negative effect closer to the obstacle, as cars were congesting more smoothly but to a greater extent.

The results of this algorithm change were both transient and often negative to the overall velocity of the system, so the change was dropped from the final algorithm shown later.  The reason for this negative effect is thought to be related to the size of field we are simulating.  In future simulations we will simulate a much larger field and so the effect of this algorithm change may become more positive as the distance from the obstacle increases.

\subsection{Lane Changing Algorithm}

The existing lane change model, introduced in section \ref{vsl}, operates by determining an advantage to be gained by changing lane and then testing if a threshold is reached by the advantage. Early incarnations of our changes to the algorithm worked to forcibly increase the advantage if  a message had been received and the vehicle was in the lane with the obstacle.  This has some positive effects, but can cause problems when the message propagates far back through the system. In the case of the message propagating beyond the reasonable extent of the need to change lane, this approach causes unnecessary congestion in the opposite lane to the obstacle that results in total congestion in a short time.

At the start of the simulation several static variables are applied to the model.  A changing threshold is applied that indicates the increased acceleration the lane change will yield; this is set by default at 0.3ms$^{-2}$ for cars in the field. The other value is the politeness factor which reduces the overall calculated advantage and which simulates the actual care drivers take when changing lane (i.e. the model may say it is advantageous to change lane, but the driver may be more polite or hesitant).

The basic algorithm operates by calculating a value of advantage (MyAdv), the disadvantage this causes to other (OthDisAdv) and then calculates whether a function of these values reaches a changing threshold. If the threshold is reached the vehicle changes lane.

\begin{equation}
\label{my_adv}
MyAdv = a_{new} - a_{old} + bias
\end{equation}
where $a_{new}$ is the acceleration in the new lane and $a_{old}$ is the acceleration in the old lane. Bias refers to a weighting to keep the vehicle in the slow lane, as operates in reality.
\begin{equation}
\label{oth_disad}
OthDisAdv = a_{behind(old)} - a_{behind(new)}
\end{equation}
where $a_{behind(old)}$ is the acceleration of the car behind in the old lane if I change lane and $a_{behind(new)}$ is the acceleration of the car behind in the new lane if I change lane.
These values are then entered into the following equation to return true or false to changing lane:
\begin{equation}
(MyAdv - p) * OthDisAdv > Thresh
\end{equation}
where $p$ is the politeness factor and $Thresh$ is the changing threshold. The form of Eq. 9 is multiplicative  so the values of MyAdv and OthDisAdv have a significant impact on each other.
In the following equations, these values are only calculated if the vehicle has been infected with a message.  If a vehicle is ignorant it will continue to use the algorithm in Eq. 9. The initial change we made was to add a value to MyAdv in Eq. 10 as such:
\begin{equation}
\label{thresh_1}
((MyAdv + V) - p) * OthDisAdv > Thresh
\end{equation}
This is very much a brute force approach and as such does not truly represent real driving in a system, where the value of V would increase as the vehicle approaches the obstacle and drop to zero after the obstacle has been passed.  This proportional addition to MyAdv is shown in Eq. 11 and Eq. 12:
\begin{gather}
Diff = \frac{Pos_{obst}}{(Pos_{me} - Pos_{obst})} \\
((MyAdv + Diff) - p) * OthDisAdv > Thresh
\end{gather}
In this adaption of the original algorithm the value $Diff$ is calculated as a function of the location of the obstacle and the vehicle's distance to it.  This value is capped at a maximum (currently 20) to prevent unrealistic behaviour (i.e. cutting in with zero safety headway), which means the effect is noticeable but quite subtle, when compared to the brute-force method in \ref{thresh_1}.  This means that as the vehicle approaches the obstacle the incentive to change lane becomes greater, reducing the appearance of congestion at the obstacle in the same lane.

\section{Simulation studies}

\begin{figure}[ht]
\centerline{\includegraphics[height=3.4cm]{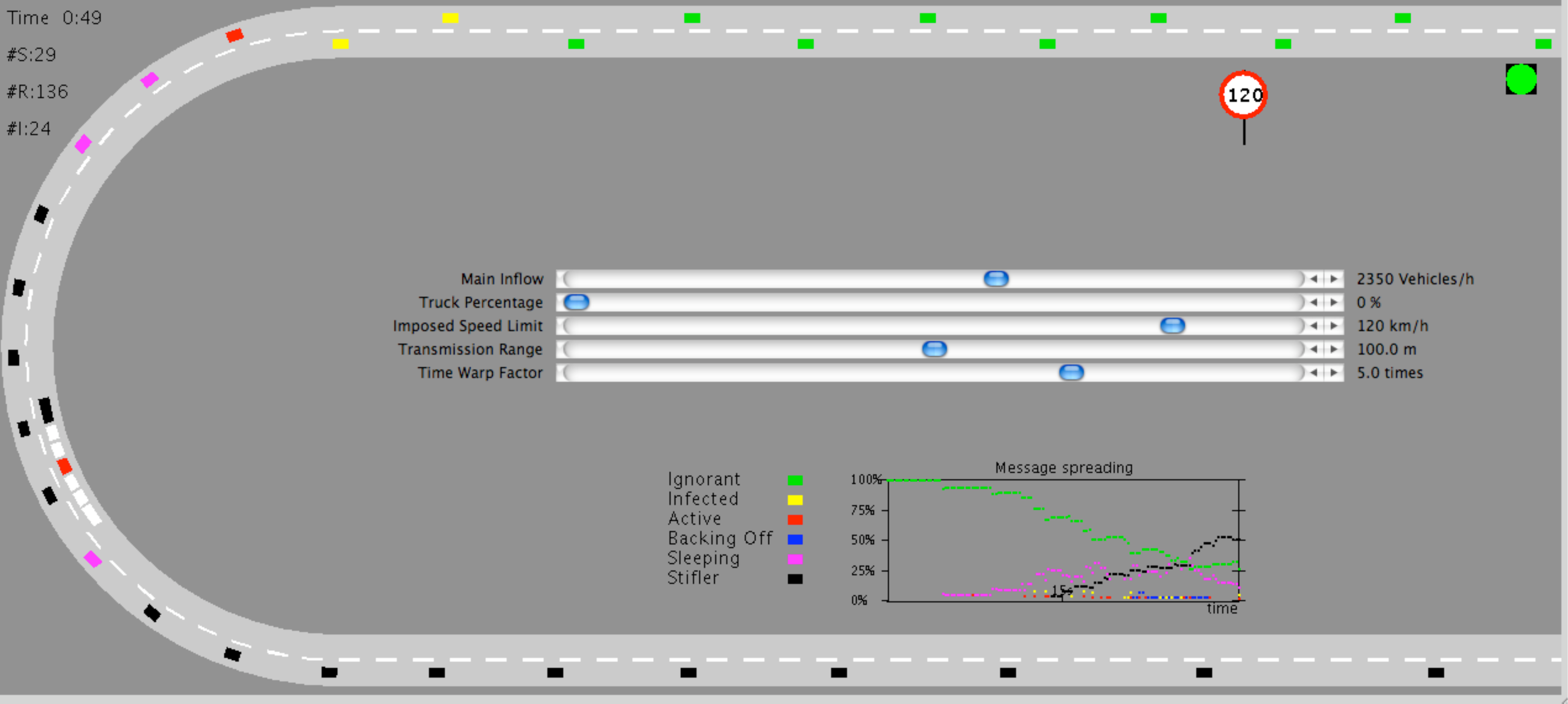}}
\caption{A graphic showing the simulator}
\end{figure}

In this section the simulation scenarios are discussed with diagrams and presentation of the results.  All the simulations were run with varying settings, so that an appreciation of all the situations that may occur (that we can control in the simulator) can be established. It is important to note, as previously mentioned, that we assume all vehicles in the simulation are equipped with the technology for message propagation.  In reality this market penetration will take many years to achieve, but many car manufacturers are working on supplying this technology soon \cite{Cars2007}.

\subsection{Velocity Experiments}

These experiments test the effectiveness of the algorithm as vehicle velocity changes, to see if the algorithms are suited to an urban (slow) or highway (fast) environment. When ignorant (i.e. have not received the message), the cars will still attempt to change lane to avoid the obstacle, but only as part of the original lane changing algorithm, and so congestion builds up in a short amount of time, for most simulations.  Below a certain network load the road will never become congested, so the traffic load of the experiment was varied for each experiment. The traffic load was also set low enough so that the algorithm can affect the flow of cars as, at high loads, this would not be possible.  To this end there is, in any system, a critical value of traffic load after which no action can prevent or reduce congestion.  

Some early simulations with low traffic loads showed that it is sometimes more efficient to be ignorant of the obstacle, and this must be taken into account, as in this case the best course of action is to drive normally, using the normal algorithm.

The following charts show results from a sample of the experiments we ran to test this theory.  By varying all the parameters we could we found that certain velocities, transmission ranges and traffic load had different effectiveness to the overall congestion in the system.  In the main the results showed that our algorithm always produced a positive effect.

\begin{figure}[ht]
\label{exit1}
\centerline{\includegraphics[height=6.2cm]{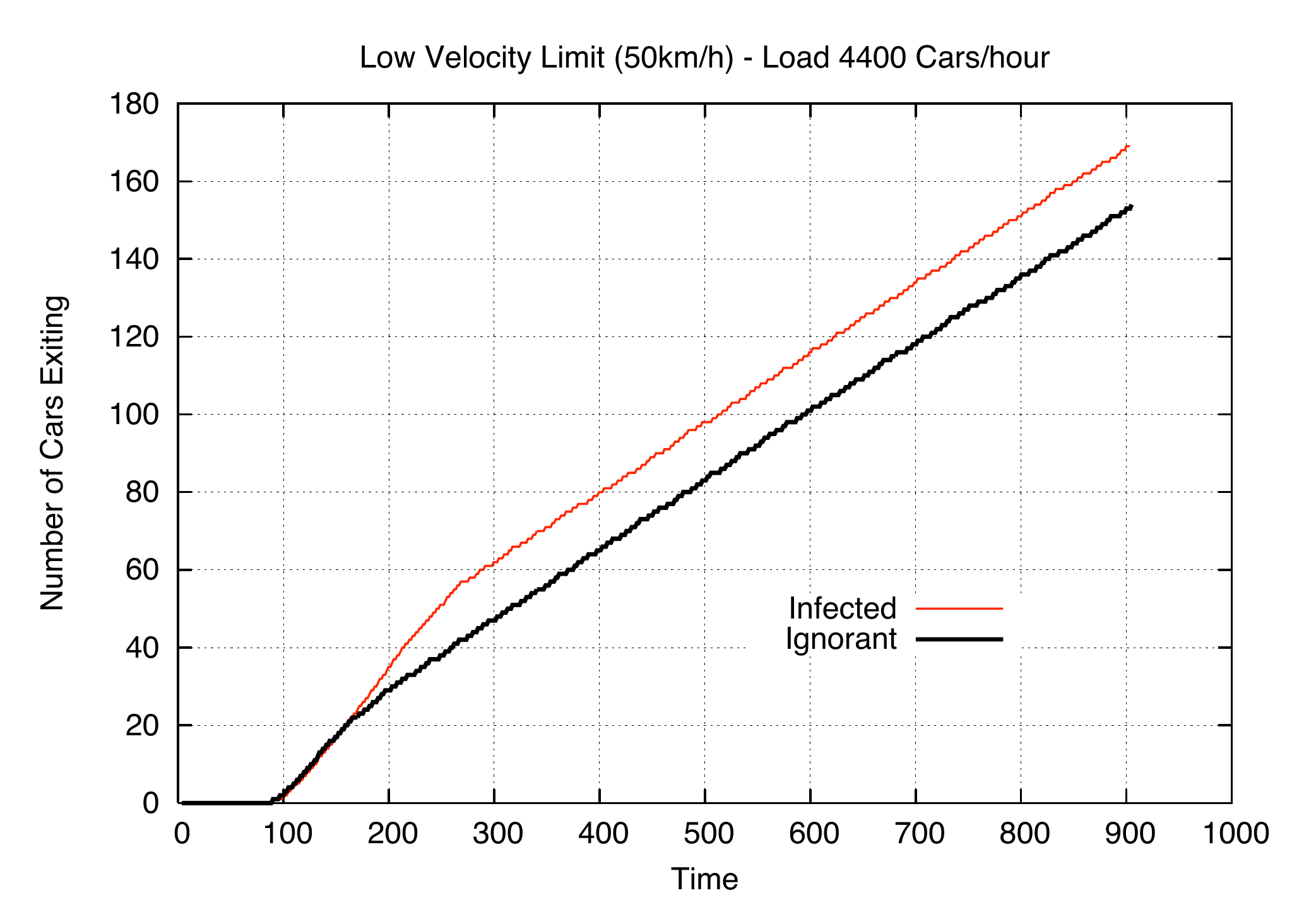}}
\caption{Comparison of exit aggregate for infected and ignorant simulations at urban velocities (below 50km/h)}
\end{figure}

\begin{figure}[ht]
\label{exit2}
\centerline{\includegraphics[height=6.2cm]{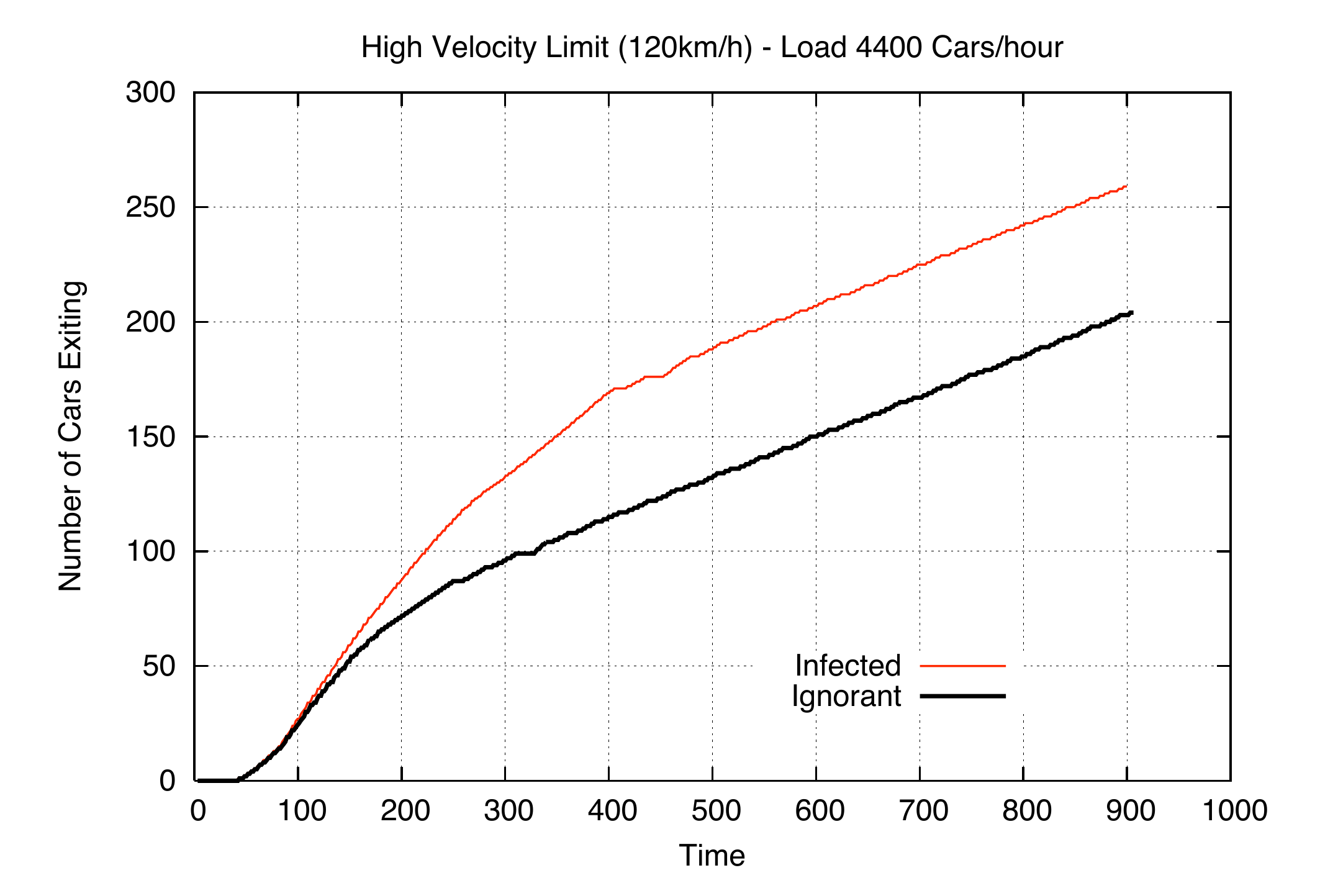}}
\caption{Comparison of exit aggregate for infected and ignorant simulations for motorway velocities}
\end{figure}

Figures Fig. 2 and Fig. 3 show the number of cars exiting the field in a simulation run as an aggregate over time.  Both show an advantage for infected cars using the advanced lane changing algorithm, but the advantage is greater at higher velocities, where the vehicles have more distance between them for the same traffic load, meaning they can more easily change lane.

We found, and Fig. 2 and Fig. 3 corroborate this, that the advanced algorithm increased the time before congestion began to build up and then once congested, the infected cars still moved through the system more efficiently.  By running the simulation for 15 minutes we can monitor the development of congestion in the system and how the flow is affected by the changed algorithm.

The results show some interesting behaviours, beyond the reduction of congestion in the system.  By analysing the first 5-7 minutes of simulation time, we can see that the development of congestion is also slower once vehicles do start to slow down.  This is because of the algorithm moving vehicles into the opposite lane to the obstacle, reducing the load on the lane with the obstacle and therefore reducing the number of stopped vehicles behind the obstacle which, when changing lane, cause a dramatic slowdown in the new lane.  This reduction in stop-and-go traffic formation is also seen elsewhere in the field when the cars are infected with the warning message and switch to our adapted algorithm.

\subsection{Position of Lane Change}

A factor that affects the build-up of congestion in the system is related to the location of the lane change.  The following results show where the lane change occurs with no communication and then using our enhanced lane change algorithm with communication active. The simulation settings were set at 4400cars/hour load, speed limit of 120km/h and a transmission range of 100m.

During the simulation the message propagates backwards towards position 0 and, with the advanced algorithm, the location of the lane-change also reduces.  When the system starts to slow and traffic becomes more dense, the lane-change moves right back, causing a slower build-up of congestion and a greater amount of free traffic, as shown by Fig. 4. This does place greater load on the opposite lane to the obstacle, but the reduction of stop-and-go behaviour negates this. In Fig. 4 the rapid reduction in position of most lane changes between 395-405 seconds and again at 475-500 seconds represents a period when the propagation of the message is continuous, and the periods of little change (of lane change position) are due to reduced propagation of the warning message. The initial peak of lane change position between 0-35 seconds represents the initialisation of the system, that cars can change lane very close to the obstacle due to the road being less loaded.

\begin{figure}[ht]
\centerline{\includegraphics[height=6.7cm]{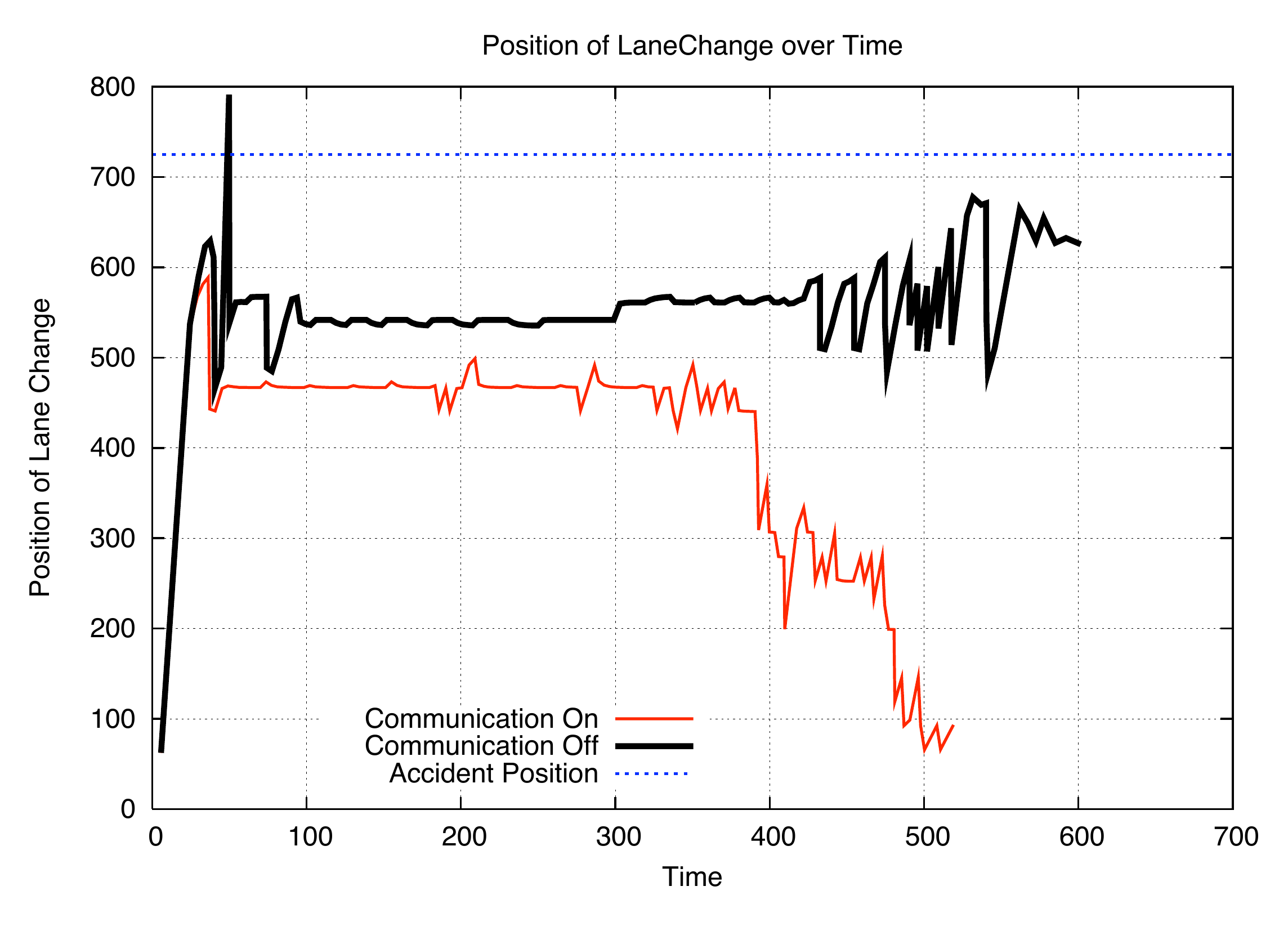}}
\caption{Chart showing location on the field of lane changes with and without communication for experiment B}
\end{figure}

\subsection{Transmission Method Experiments}

In this experiment the available transmission methods are tested with the new algorithm, to see how they affect the overall congestion in the system. The following figure shows a simulation run until congestion is present at the origin of the field (i.e. position = 0).  The simulation is stopped when the congestion reaches the origin as after this point the algorithm is not affecting the traffic.  The values represent the proportion of cars leaving the field in relation to the cars entering $\left( \frac{Cars Exiting}{Cars Arriving}\right)$.  This indicates how vehicles are flowing through the system, where an increase of the gradient represents free flow and a decrease represents congestion.

The models shown in Fig. 5 are simple flooding, where the message is rebroadcast just once, edge detection which is explained in section $III-A$ and distance detection, which is a different probabilistic method and a mixture of edge and distance detection.  These models are all running in the simulation, but the mixed offers the best simulation of a real epidemic protocol.

As can be seen from Fig. 5 the edge detection method alone offers little improvement over no propagation, and the simple flooding and distance detection methods offer a good initial advantage (0-200 seconds) but then suffer very fast congestion build-up.  The mixture of edge and distance detection algorithm, with our changes to the lane change model offers excellent results keeping near free flow until approx. 420 seconds, when the network then slows and starts to congest, but this takes longer (approx. 250 seconds from the first slowdown) than the other algorithms.

\begin{figure}[ht]
\centerline{\includegraphics[height=6.2cm]{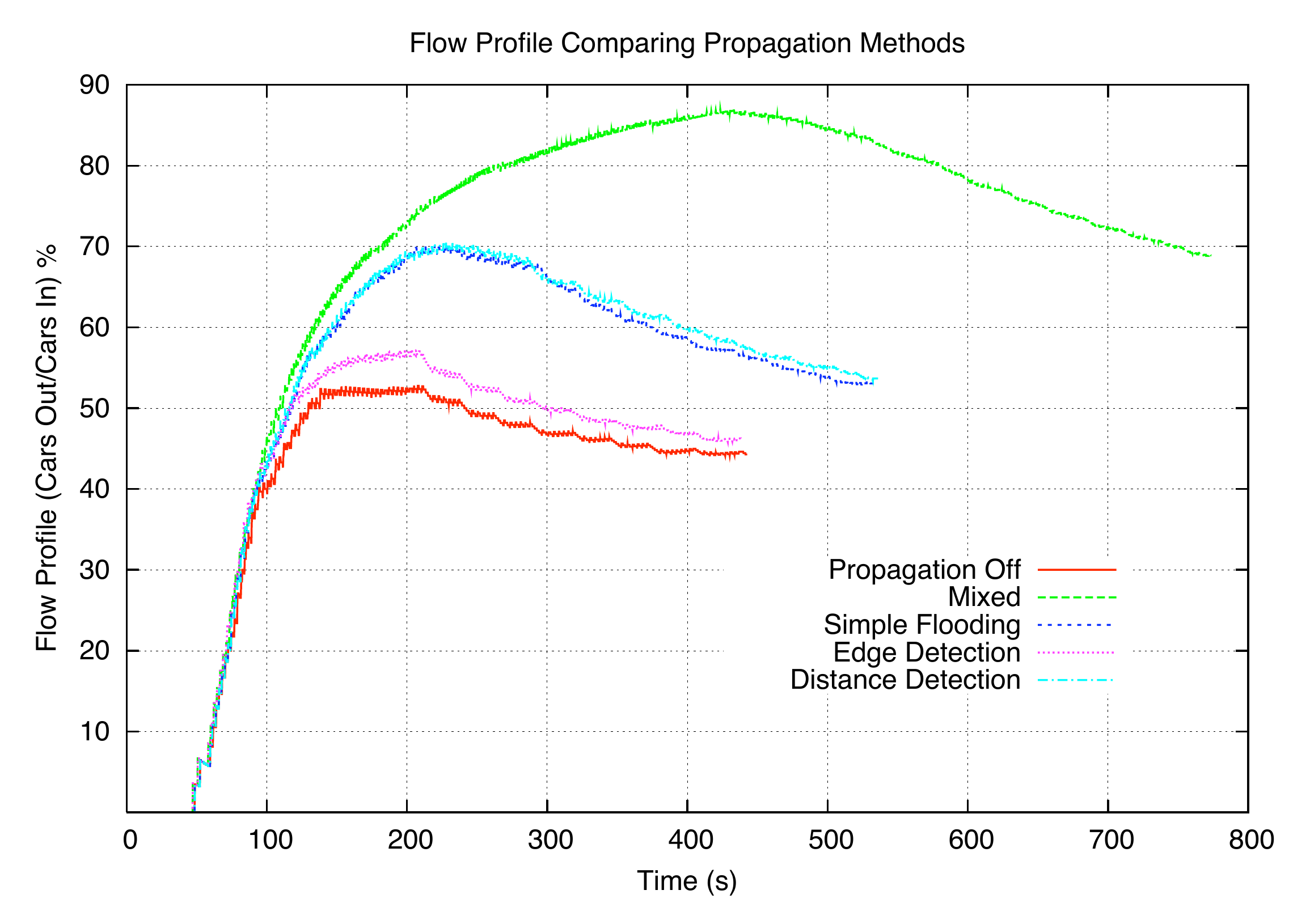}}
\caption{Comparison of message propagation methodologies in experiment C}
\end{figure}

The addition of this propagation method and the changed algorithm prevents several congestion-causing situations to occur.  The main situation avoided is that vehicles are unable to change lane to avoid the obstacle and begin to slow down, but then do change lane causing the cars behind to slow.  This cause has been seen to initiate the build-up of congestion and by earlier warning of the obstacle the cars can change lane at a high velocity. Another factor of the system is that when the congestion is initiated, the cars will fill up behind the obstacle, unable to change lane.  With the adapted algorithm the extra incentive to change lane means that the opposite lane fills first and so traffic can still move, increasing the time before the whole system becomes congested.

\subsection{System Velocity}

The average velocity through a system is of great importance.  If a higher average velocity can be achieved the throughput will be higher than if there is much slowing of traffic. The following two figures show the average velocity calculated for intervals of 10 metres on the $x$ axis and an interval of 30 seconds across the $y$ axis.  Each point represents the average velocity at that time/position interval.  In both figures there is a noticeable slowdown as the vehicles pass the obstacle.  This can be accounted for by the IDM attempting to retain a minimum safe distance between vehicles.

\begin{figure}[ht]
\centerline{\includegraphics[height=6.7cm]{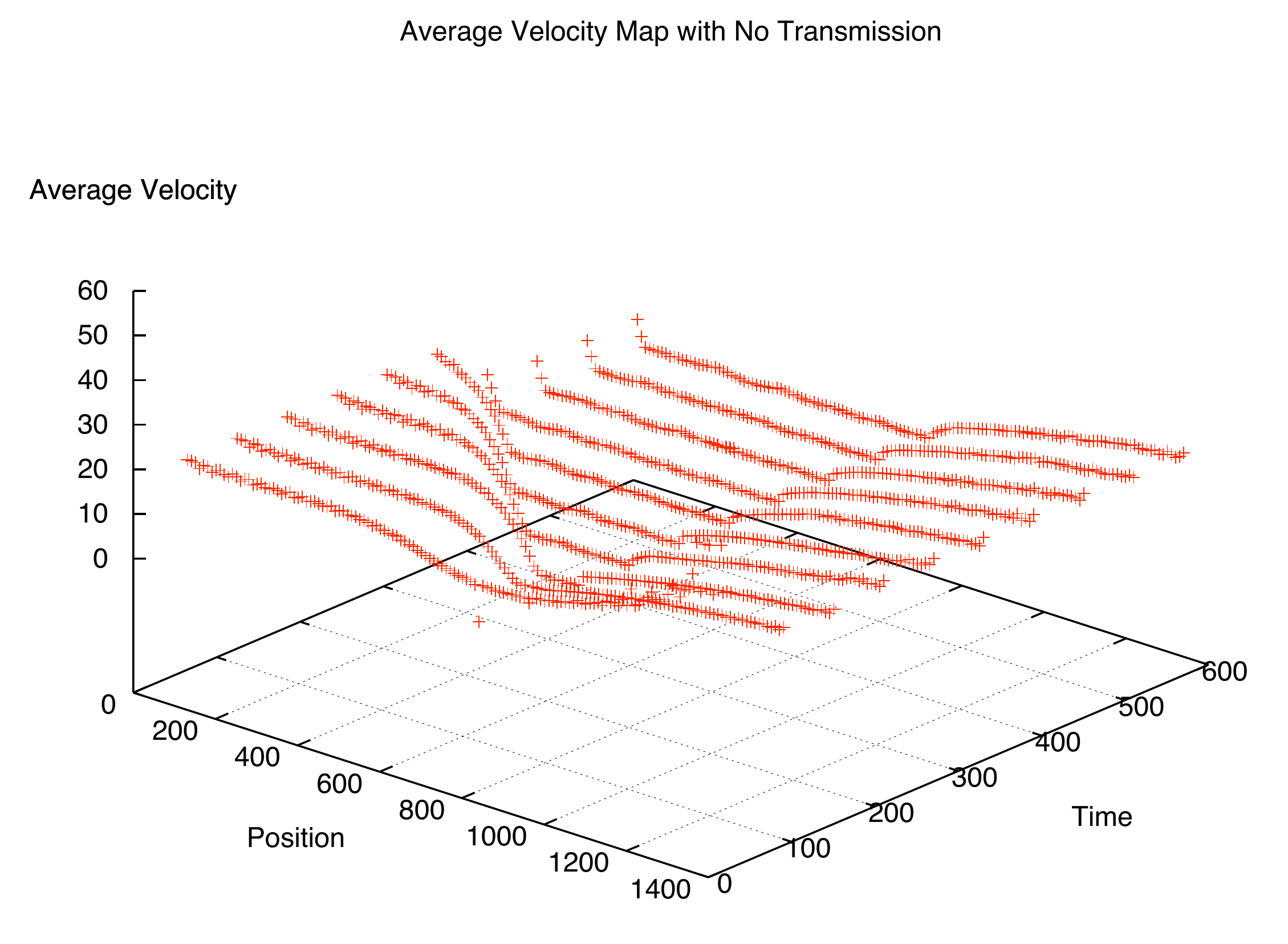}}
\caption{Average velocities across the system without transmission}
\end{figure}

\begin{figure}[ht]
\centerline{\includegraphics[height=6.7cm]{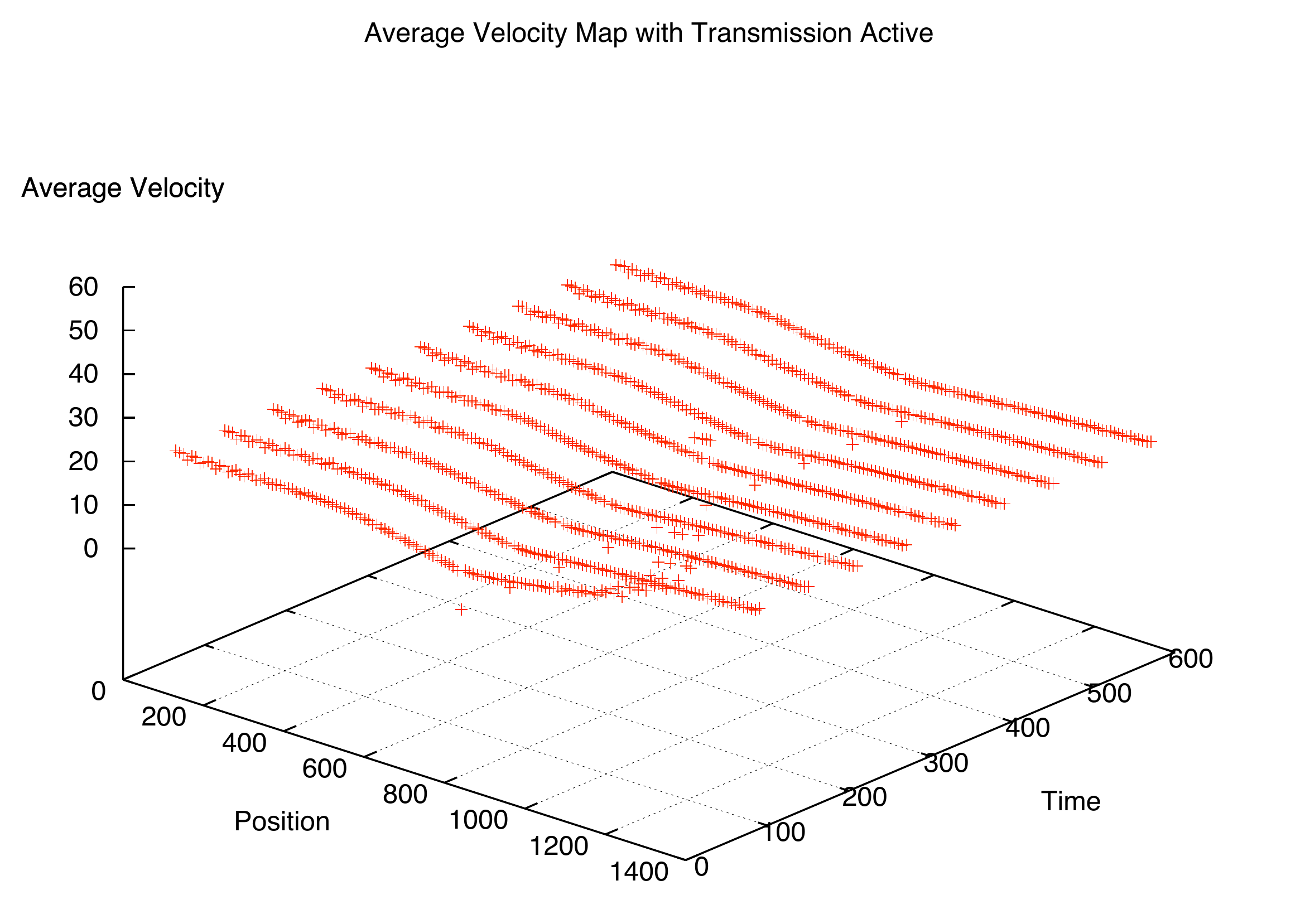}}
\caption{Average velocities across the system with active transmission}
\end{figure}

Fig.6 shows that after an initial even velocity through the field, congestion begins to build up at the position of the obstacle between 100-200 seconds, which causes a slowdown further back to position 0.  By time 330 seconds the congestion has reached position 0 and the average velocity falls from 20-25ms$^{-1}$ to 0-5ms$^{-1}$.

As can be seen in Fig.7, there is a uniform average velocity before and after the obstacle during the whole period of the simulation (10 minutes).  This enforces the results from the other simulations and proves there is a better flow of traffic through the network, as well as a reduction in the build-up of congestion, when effective transmission of the road condition occurs.

We note for both figures there is a spike in velocities (between time 0-10 and position 1200-1400). This is the period when the first cars are leaving the field.  As they have no cars in front they can accelerate up to the full speed limit unhindered. To remove this artefact future simulations will have a 'warm-up' period, with low traffic load, that initialises the field.

\addvspace{25pt}
\section{Conclusions}

This paper has presented a simulation of a specific road condition, that of an accident blocking a lane on a dual carriageway.  The simulation uses the coupled approach to mobility modelling and network simulation in a single process.  The models and algorithms that represent the vehicles and network traffic have been implemented according to well-known and standardised models.  We have shown that by adapting the algorithms when information about the accident is received via wireless transmission, we can reduce the build-up of congestion near the obstacle and improve the network throughput, both in terms of quantity of cars and average velocity.

The various algorithms we tested achieved overall improvement in the majority of cases. In some simulations the improvement was not only in the prevention of congestion overall, but by also keeping a consistent average velocity through the network, which helps to reduce the effects of stop-and-go traffic and smooth
possible congestion 'waves' that emanate from the source backwards. The driver model and lane changing algorithms come from highly validated sources, and so the adaptions we have made are highly realistic and can show the effect of even simple changes (as in the brute force addition of a value to MyAdv in section $IV - C$). In order to fully test our algorithms we ran numerous tests with a wide variety of parameters, in order to test for any transient or artifactual effects.  From these repetitive runs we established that most effects were long-lasting and that where those were transient, this was only due to a traffic load that was impossible to prevent congestion (as corroborated by control tests). 

In the common situation where the obstacle is temporary (i.e. a vehicle malfunction), any reduction in congestion build-up allows for the obstacle to be removed, before the velocity of all vehicles behind the obstacle drops to zero. In more complex roadways the advanced warning could also lead to a change of route, so drivers can avoid the section of road where the obstacle exists. 

In the size of field simulated here we are between the microscopic and macroscopic scale of simulation, which is achieved seamlessly by the use of a coupled model of simulation. The tool we performed the simulations with was lightweight and so we could easily implement new algorithms and protocols.  In order to run more complex networks we would require a more complex simulation engine for vehicle traffic. With this increase in complexity the field size will increase and therefore a more powerful network simulator is required. The proof of concept that this paper provides will lead to further work in this field, including the use of parallel and distributed computational resource.

\addvspace{25pt}
Maziar Nekovee acknowledges the Royal Society for supporting his work through an Industry Fellowship.

\bibliographystyle{ieeetr}
\bibliography{../../References}

\begin{thebibliography}{10}

\bibitem{Torrent-Moreno2007}
M.~Torrent-Moreno, {\em Inter Vehicle Communications, Acheiving Safety in a
  Distributed Wireless Environment: Challenges, Systems and Protocols}.
\newblock PhD thesis, Universit{\"a}t Karlsruhe, 2007.

\bibitem{1682943}
M.~Nekovee, ``Modeling the spread of worm epidemics in vehicular ad hoc
  networks,'' {\em Vehicular Technology Conference, 2006. VTC 2006-Spring. IEEE
  63rd}, vol.~2, pp.~841--845, 7-10 May 2006.

\bibitem{treiber-2000-62}
M.~Treiber, A.~Hennecke, and D.~Helbing, ``Congested traffic states in
  empirical observations and microscopic simulations,'' {\em Physical Review
  E}, vol.~62, p.~1805, 2000.

\bibitem{Treiber1999}
A.~Kesting, M.~Treiber, and D.~Helbing, ``General lane-changing model {MOBIL}
  for car-following models,'' {\em Traffic Flow Theory 2007}, pp.~pp 86--94,
  1999.

\bibitem{4212940}
M.~Nekovee and B.~Bogason, ``Reliable and effcient information dissemination in
  intermittently connected vehicular adhoc networks,'' {\em Vehicular
  Technology Conference, 2007. VTC2007-Spring. IEEE 65th}, pp.~2486--2490,
  22-25 April 2007.

\bibitem{4248378}
IEEE, ``{IEEE} standard for information technology-telecommunications and
  information exchange between systems-local and metropolitan area
  networks-specific requirements - part 11: Wireless {LAN} medium access
  control ({MAC}) and physical layer ({PHY}) specifications,'' {\em IEEE Std
  802.11-2007 (Revision of IEEE Std 802.11-1999)}, pp.~C1--1184, June 12 2007.

\bibitem{weinmiller96analyzing}
J.~Weinmiller, H.~Woesner, and A.~Wolisz, ``Analyzing and improving the {IEEE}
  802.11-{MAC} protocol for wireless {LAN}s,'' in {\em Proceedings of the 4th
  International Workshop on Modeling, Analysis, and Simulation of Computer and
  Telecommunication Systems ({MASCOTS} '96)}, pp.~200--206, 1996.

\bibitem{Yin2003}
J.~Yin, T.~ElBatt, G.~Yeung, B.~Ryu, S.~Habermas, H.~Krishn, and T.~Talty,
  ``Performance evaluation of safety applications over {DSRC} vehicular ad hoc
  networks,'' {\em VANET '04}, 2003.

\bibitem{Struzak2006}
R.~Struzak, ``Radio-wave propagation basics,'' tech. rep., ICTP-ITU-URSI School
  on Wireless Networking for Development, 2006.

\bibitem{Laasonen2003}
K.~Laasonen, ``Radio propagation modeling,'' tech. rep., University of
  Helsinki, 2003.

\bibitem{Torrent-Moreno2007a}
M.~Torrent-Moreno, ``Inter-vehicle communications: Assessing information
  dissemination under safety constraints,'' tech. rep., Institute of
  Telematics, University of Karlsruhe, Germany, 2007.

\bibitem{Cars2007}
D.~Waters, ``Connected cars 'promise safer roads','' {\em BBC News 2007}, 2007.

\end{thebibliography}

\end{document}